\documentstyle [12pt] {article}

\newcommand{\nwc}{\newcommand}
\nwc{\cl}  {\clubsuit}

\nwc{\hyp} {\hyphenation}
\nwc{\be}  {\begin{equation}}
\nwc{\ee}  {\end{equation}}
\nwc{\ba}  {\begin{array}}
\nwc{\ea}  {\end{array}}
\nwc{\bdm} {\begin{displaymath}}
\nwc{\edm} {\end{displaymath}}
\nwc{\bea} {\be\ba{rcl}}
\nwc{\eea} {\ea\ee}
\nwc{\ben} {\begin{eqnarray}}
\nwc{\een} {\end{eqnarray}}
\nwc{\bda} {\bdm\ba{lcl}}
\nwc{\eda} {\ea\edm}
\nwc{\bc}  {\begin{center}}
\nwc{\ec}  {\end{center}}
\nwc{\ds}  {\displaystyle}
\nwc{\bmat}{\left(\ba}
\nwc{\emat}{\ea\right)}
\nwc{\non} {\nonumber}
\nwc{\bib} {\bibitem}
\nwc{\lra} {\longrightarrow}
\nwc{\Llra}{\Longleftrightarrow}
\nwc{\ra}  {\rightarrow}
\nwc{\Ra}  {\Rightarrow}
\nwc{\lmt} {\longmapsto}
\nwc{\prl} {\partial}
\nwc{\iy}  {\infty}
\nwc{\ol}  {\overline}
\nwc{\hm}  {\hspace{3mm}}
\nwc{\lf}  {\left}
\nwc{\ri}  {\right}
\nwc{\lm}  {\limits}
\nwc{\lb}  {\lbrack}
\nwc{\rb}  {\rbrack}
\nwc{\ov}  {\over}
\nwc{\pr}  {\prime}
\nwc{\nnn} {\nonumber \vspace{.2cm} \\ }
\nwc{\Sc}  {{\cal S}}
\nwc{\Lc}  {{\cal L}}
\nwc{\Rc}  {{\cal R}}
\nwc{\Dc}  {{\cal D}}
\nwc{\Oc}  {{\cal O}}
\nwc{\Cc}  {{\cal C}}
\nwc{\Pc}  {{\cal P}}
\nwc{\Mc}  {{\cal M}}
\nwc{\Ec}  {{\cal E}}
\nwc{\Fc}  {{\cal F}}
\nwc{\Hc}  {{\cal H}}
\nwc{\Kc}  {{\cal K}}
\nwc{\Xc}  {{\cal X}}
\nwc{\Gc}  {{\cal G}}
\nwc{\Zc}  {{\cal Z}}
\nwc{\Nc}  {{\cal N}}
\nwc{\fca} {{\cal f}}
\nwc{\xc}  {{\cal x}}
\nwc{\Ac}  {{\cal A}}
\nwc{\Bc}  {{\cal B}}
\nwc{\Uc}  {{\cal U}}
\nwc{\Vc}  {{\cal V}}
\nwc{\Th} {\Theta}
\nwc{\th} {\theta}
\nwc{\vth} {\vartheta}
\nwc{\eps}{\epsilon}
\nwc{\si} {\sigma}
\nwc{\Gm} {\Gamma}
\nwc{\gm} {\gamma}
\nwc{\bt} {\beta}
\nwc{\La} {\Lambda}
\nwc{\la} {\lambda}
\nwc{\om} {\omega}
\nwc{\Om} {\Omega}
\nwc{\dt} {\delta}
\nwc{\Si} {\Sigma}
\nwc{\Dt} {\Delta}
\nwc{\al} {\alpha}
\nwc{\vph}{\varphi}
\nwc{\zt} {\zeta}
\def\tr{\mathop{\rm tr}}
\def\Tr{\mathop{\rm Tr}}

\def\VEV#1{\left\langle #1\right\rangle}

\def\abs#1{\left| #1\right|}
\def\pr#1{#1^\prime}

\nwc{\Id}  {{\bf 1}}
\nwc{\diag} {{\rm diag}}
\nwc{\inv}  {{\rm inv}}
\nwc{\mod}  {{\rm mod}}
\nwc{\hal} {\frac{1}{2}}
\nwc{\tpi}  {2\pi i}

\def\pr#1{Phys. Rev. {\bf #1}}

\def\NP#1{Nucl. Phys.~{\bf #1}}
\def\NC#1{Nuovo Cim.~{\bf #1}}
\def\PH#1{Physica {\bf #1}}
\def\PL#1{Phys. Lett.~{\bf #1}}
\def\PR#1{Phys. Rev.~{\bf #1}}
\def\PRP#1{Phys. Rep.~{\bf #1}}

\def\RMP#1{Rev. Mod. Phys.~{\bf #1}}

\def\ZP#1{Z. Phys.~{\bf #1}}

\def\MeV {\,{\rm  MeV}}

\newsavebox{\nnin} \sbox{\nnin}{$\hspace{1mm}\in\kern -.8em /
                   \hspace{1mm}$}

\newcommand{\sub}{\subset}
\newsavebox{\nnsub} \sbox{\nnsub}{$\hspace{1mm}\sub\kern -.9em /
            \hspace{1mm}$}

\def\KK{{\rm I\kern -.2em  K}}
\def\NN{{\rm I\kern -.16em N}}
\def\RR{{\rm I\kern -.2em  R}}
\def\ZZ{Z \kern -.43em Z}
\def\QQ{{\rm \kern .25em
             \vrule height1.4ex depth-.12ex width.06em\kern-.31em Q}}
\def\CC{{\rm \kern .25em
             \vrule height1.4ex depth-.12ex width.06em\kern-.31em C}}
\def\ZZZ{Z\kern -0.31em Z}

\newcounter{app}
\def\app{\par
 \addtocounter{app}{1}
 \def\thesection{\Alph{app}}
 \def\ksection{\Alph{app}}}

\def\appendix#1{\app\sect{#1}}

\newcommand{\sect}[1]{ \section{#1} \setcounter{equation}{0} }

\begin{document}

\begin{titlepage}

  \title{The linear meson model\\ and\\ chiral perturbation
    theory\thanks{Supported by the Deutsche Forschungsgemeinschaft}}

\author{{\sc D.--U. Jungnickel\thanks{Email: 
D.Jungnickel@thphys.uni-heidelberg.de}} \\
 \\ and \\ \\
{\sc C. Wetterich\thanks{Email: C.Wetterich@thphys.uni-heidelberg.de}} 
\\ \\ \\
{\em Institut f\"ur Theoretische Physik} \\
{\em Universit\"at Heidelberg} \\
{\em Philosophenweg 16} \\
{\em 69120 Heidelberg, Germany}}

\date{}
\maketitle

\begin{picture}(5,2.5)(-350,-450)
\put(12,-114){HD--THEP--97--9}
\put(12,-131){\today}
\end{picture}

\thispagestyle{empty}

\begin{abstract}
  We compare the linear meson model and chiral perturbation theory in
  next to leading order in the quark mass expansion. In particular, we
  compute the couplings $L_4$--$L_8$ of chiral perturbation theory as
  functions of the parameters of the linear model. They are induced by
  the exchange of $0^{++}$ scalar mesons. We use a phenomenological
  analysis of the effective vertices of the linear model in terms of
  pseudoscalar meson masses and decay constants. Our results for the
  $L_i$ agree with previous phenomenological estimates.
\end{abstract}

\end{titlepage}

\sect{Introduction}
\label{Introduction}

Spontaneously broken chiral symmetry is a basic ingredient for the
description of light mesons in the context of QCD. It is based on the
observation that in the limit of vanishing light quark masses,
$m_u,m_d,m_s\ra0$, the QCD Lagrangian exhibits a chiral $SU_L(3)\times
SU_R(3)$ flavor invariance. Since in this limit the light meson
spectrum only shows an explicit vector--like $SU_V(3)$ symmetry the
full chiral flavor invariance must be broken spontaneously by a
``chiral condensate'`. For small quark masses $m_q$ ($q=u,d,s$ here)
one expects the existence of some sort of systematic
expansion\footnote{Not all quantities are analytic in $m_q$, though.
  For instance, the pion mass is proportional to $(m_u+m_d)^{1/2}$.
  Further non--analyticities in higher couplings are induced by
  quantum fluctuations of massless Goldstone bosons for $m_q\ra0$.} in
powers of $m_q$. To a given order in $m_q$ the spontaneously broken
chiral symmetry only allows for a finite number of effective couplings
entering the description of the pseudoscalar octet ($\pi,K,\eta$). To
lowest order these are the pion decay constant $F_0$ and the
proportionality constant $B_0$ between the squared meson masses and
the quark masses, e.g., $M_{\pi^\pm}^2=B_0(m_u+m_d)$.  With increasing
order in $m_q$ the number of independent chiral invariants and
therefore of free parameters grows rapidly. Nevertheless, useful
information can be extracted from the low orders of the quark mass
expansion if $m_q$ is sufficiently small.

There are different ways of realizing the symmetry content of QCD
within low energy models for the light mesons. The minimal model is
the nonlinear sigma model for the pseudoscalar octet which is
described by a special unitary $3\times3$ matrix $\tilde{U}$,
$\tilde{U}^\dagger\tilde{U}=1$, $\det\tilde{U}=1$
\cite{Wei79-1,GL82-1}.  It can be extended to include the
$\eta^\prime$ meson explicitly if the constraint $\det\tilde{U}=1$ is
dropped. In the linear sigma model \cite{GML60-1} the low lying scalar
and pseudoscalar mesons are grouped into a field $\Phi$ transforming
in the $(\ol{\bf 3},{\bf 3})$ representation of the chiral group
\cite{CH73-1,JW96-2}. Here the additional $0^{++}$ octet describes the
scalar mesons ($a_0,K_0^*,f_0$) and the $0^{++}$ singlet corresponds
to the ``$\sigma$--resonance''. Further extensions of mesonic models
also include fields for the lightest vector and pseudovector mesons
(see, e.g., \cite{Mei88-1,JW96-2} and references therein). A
projection of these models onto the pseudoscalar octet states can be
computed by first integrating the mesonic quantum fluctuations and
subsequently accounting for the exchange of the $\eta^\prime$, the
scalar, vector and pseudovector mesons (depending on the model) by
solving their field equations.  Technically, all masses and vertices
are contained in the {\em effective action}\footnote{Often the terms
  ``effective model'' or ``effective action'' are also used to
  describe the result of a reduction of degrees of freedom. In this
  case we will rather employ here the term ``low energy model'' and
  reserve the term ``effective action'' for the generating functional
  of $1PI$ Green functions.} for the mesons which is the generating
functional for the $1PI$ Green functions. All quantum fluctuations are
already included in the effective action. (Of course, the concept of
one--particle--irreducibility depends on the particle content of the
model. For example, a four pion interaction mediated by the exchange
of a $\rho$--meson becomes $1PI$ in the pure nonlinear or linear sigma
model.)  As far as only the pseudoscalar octet (or nonet) is concerned
a common language for the different mesonic models can be found by
computing the effective action restricted to the corresponding fields.
Formally, the latter can be obtained from the effective action of the
linear model by solving the field equations for the scalar fields as
functions of the pseudoscalar fields and inserting the result into the
effective action. (A similar procedure can be applied to models which
contain fields for the vector and axial--vector mesons.) We emphasize
that the symmetry content of all these mesonic models is the same as
far as the pseudoscalar octet is concerned.  All relations that follow
from spontaneously broken chiral symmetry if the masses and
interactions of the pseudoscalars are computed to a given order in
$m_q$ are identical. This also holds if the quark mass expansion is
connected to a simultaneous expansion in powers of derivatives for the
pseudoscalar fields as usually done in chiral perturbation theory.
The quark mass expansion is, however, not unique. Instead of
truncating the computation of physical quantities at a given order in
$m_q$ one may alternatively choose to include in the effective action
only those invariants which contribute to a given order in $m_q$. If
no further truncations are performed the physical observables
calculated from these invariants contain contributions which are
formally of higher order in $m_q$. This procedure has been followed in
\cite{JW96-2} for the linear meson model. It amounts to a partial
resummation of the quark mass expansion for physical quantities. The
results obtained from the linear and the nonlinear model may therefore
differ by higher order quark mass effects.  Beyond pure symmetry
relations the different models typically contain additional
assumptions which may lead to further predictive power for mesonic
observables.

Chiral perturbation theory describes the ``Goldstone degrees of
freedom'' by a nonlinear sigma model. It is valid for momenta
sufficiently below the $\rho$--meson and $\sigma$--resonance masses.
There the parameters $F_0$, $B_0$ as well as the couplings $L_i$ which
appear at next to leading order in the quark mass expansion are
specified at some normalization scale $\mu$. The propagators and
vertices of the effective action or the physical observables are then
computed using a loop expansion which accounts for the fluctuations of
the pseudoscalar mesons. This expansion apparently converges if the
momenta and $\mu$ are not too large \cite{GL82-1}.  The most important
implicit assumption is simply that higher orders in the quark mass
expansion can be neglected. Since this is the minimal model pure
symmetry relations are most easily visualized in this context.

Within the linear meson model one may assume that a description in
terms of the field $\Phi$ and quark degrees of freedom holds up to a
``compositeness scale'' $k_\Phi$ around $(600-700)\MeV$. (In addition,
one may use explicit vector fields.) If, furthermore, one assumes that
the Yukawa coupling between $\Phi$ and the quarks is sufficiently
strong at the scale $k_\Phi$ this model turns out to be very
predictive for the effective mesonic action. This is a consequence of
a fast evolution towards partial infrared fixed points of its running
couplings \cite{JW96-1}. We will follow here a more modest
``phenomenological'' approach \cite{JW96-2} and concentrate only on
properties of the effective action for the linear meson model. The
main assumption made here are the ones of reference \cite{JW96-2}:
\begin{description}
\item[(i)] Terms with more than two derivatives can be neglected in
  the effective action if the latter is expanded in powers of
  differences of momenta from an appropriately chosen average momentum
  of the corresponding $SU_V(3)$ multiplet, like
  $q_0^2=-(2M_{K^\pm}^2+M_{\pi^\pm}^2)/3$ for the pseudoscalar octet.
  This assumption should hold with the exception of contributions from
  the (tree--)exchange of vector or axial--vector fields. It is
  substantiated by the apparent smallness of the momentum dependence
  of effective couplings which is induced by meson loops
  \cite{JW96-2}. Furthermore, we neglect (as in \cite{JW96-2}) certain
  higher order invariants with two derivatives involving high powers
  of the chiral condensate $\sigma_0$.
\item[(ii)] The explicit flavor symmetry breaking by current quark
  masses appears in the linear meson model only in form of a {\em
    linear} source term
  \begin{equation}
    \label{SourceTerm}
    \Lc_j=-\hal\Tr
    \left(\Phi^\dagger\jmath+\jmath^\dagger\Phi\right)
  \end{equation}
  \begin{equation}
    \label{Current}
    \jmath=\jmath^\dagger=a_q M_q
    \equiv a_q\left(
    \begin{array}{rcl}
      m_u & 0   & 0   \\
      0   & m_d & 0   \\
      0   & 0   & m_s 
    \end{array}\right) \; .
  \end{equation}
  This is motivated by the way how mesons arise as quark--antiquark
  composite states at the scale $k_\Phi$ \cite{EW94-1,JW96-4}.
  Deviations from this assumption arise from current quark mass
  dependencies of quark loop contributions to the effective
  four--quark vertex at the compositeness scale. Since typical momenta
  are larger or of the order of $k_\Phi$ these effects are small.
\item[(iii)] Invariants in the effective action which contribute only
  beyond quadratic order in the quark mass expansion can be neglected
  in the phenomenological analysis.  We emphasize here that the quark
  mass expansion apparently converges faster in the linear than in the
  nonlinear model \cite{JW96-2}. This is related to effective
  resummations which are particularly important in the case of large
  mixing of states.
\end{description}
It is important to note that our assumptions do not restrict the
effective action of the linear meson model to renormalizable
interactions. The appearance of ``non--renormalizable interactions'',
i.e. couplings with negative mass dimension, is, of course, not only
possible but even necessary for the generating functional of $1PI$
Green functions of the linear meson model. It is, however, not clear a
priori how large the corresponding couplings are in units of the
chiral condensate $\sigma_0$. It was found in \cite{JW96-2,JW96-3}
that such non--renormalizable interactions play a crucial role for the
compatibility of this model with phenomenology.  They can be explained
to a large extent by mixing with or exchange of higher mass states.

The purpose of this work is a comparison of the effective action of
chiral perturbation theory within the framework of the nonlinear sigma
model on one side with that of the linear meson model on the other
side.  Naturally, such a comparison must be restricted to the
pseudoscalar fields as the only degrees of freedom present in the
nonlinear model. Both models have free parameters which can be fixed
by using experimental input. As far as the same quantities are used as
input --- the meson masses $M_{\pi^\pm}$, $M_{K^\pm}$ and $M_{\eta}$
as well as the decay constants $f_\pi$ and $f_K$ --- and only symmetry
relations are employed there is, of course, no difference between the
models.  In the linear model we also use $M_{\eta^\prime}$ and the
decay constants $f_\eta$ and $f_{\eta^\prime}$ as obtained from the
decays $\eta\ra2\gamma$, $\eta^\prime\ra2\gamma$ (related to
singlet--octet mixing) for a determination\footnote{Actually, only two
  of the three observables $f_\eta$, $f_{\eta^\prime}$ and $M_\eta$
  are needed as input whereas the third one comes out as a rather
  successful ``prediction'' of the linear model \cite{JW96-2}.} of the
parameters \cite{JW96-2} relevant for the present work. From this
information we determine the couplings $L_i$ which appear in next to
leading order of chiral perturbation theory.  More precisely, we will
compute within the linear model effective couplings $\tilde{L}_i$
which multiply proper vertices with the same structure as those
appearing as interactions in next to leading order in chiral
perturbation theory.  In the linear model the effective derivative
terms are evaluated for momenta corresponding to the location of the
pole for an average pseudoscalar octet mass
$q_0^2=-(2M_{K^\pm}^2+M_{\pi^\pm}^2)/3$. For a full comparison one
should evaluate in chiral perturbation theory the same effective
vertices. For simplicity we will identify here the $\tilde{L}_i$ with
the couplings $L_i(\mu)$ of chiral perturbation theory, evaluated at a
renormalization scale $\mu^2=-q_0^2$. This minimizes the logarithms in
the loop expansion of chiral perturbation theory.  Nevertheless, for a
quantitatively precise comparison the same $1PI$ vertices would have
to be computed in the framework of chiral perturbation theory.

Recent estimates of the current quark masses in both the nonlinear
\cite{Leu96-1} and the linear \cite{JW96-3} sigma model show very good
agreement. The estimate of the higher order couplings $L_i$ presented
here confirms this general picture. Within errors, the results of the
linear model are compatible with the estimates from chiral
perturbation theory. For $\mu=400\MeV$ the couplings $\tilde{L}_5$ and
$\tilde{L}_8$ of the linear meson model turn out to be somewhat
smaller than the central values of the couplings $L_5(\mu)$ and
$L_8(\mu)$, respectively, in chiral perturbation theory. This may
partly be due to the uncertainties in the optimal value for $\mu$ or
the loop effects which are neglected in our comparison.  On the other
hand, it was observed earlier \cite{GL82-1,JW96-2} that the quark mass
expansion converges relatively slowly for the nonlinear model. It
seems therefore also conceivable that the observed differences reflect
higher order quark mass corrections which are neglected in next to
leading order chiral perturbation theory but are included in our
analysis of the linear model.

In the linear sigma model the scalar and pseudoscalar meson fields are
represented by a complex $3\times3$ matrix $\Phi$. The effective
potential can be constructed from the invariants
\begin{equation}
  \label{TTT01}
  \begin{array}{rclcrcl}
    \ds{\rho} &=& \ds{\Tr\Phi^\dagger\Phi} &,&
    \ds{\xi} &=& \ds{\det\Phi+\det\Phi^\dagger}\nnn
    \ds{\tau_2} &=& \ds{\frac{3}{2}
      \Tr\left(\Phi^\dagger\Phi-\frac{1}{3}\rho\right)^2} &,&
    \ds{\tau_3}&=& \ds{
      \Tr\left(\Phi^\dagger\Phi-\frac{1}{3}\rho\right)^2}\; .
  \end{array}
\end{equation}
We use a polynomial expansion around their values for the
expectation value of $\Phi$ in presence of equal light current quark
masses,
\begin{equation}
  \label{TTT02}
  \VEV{\Phi}=\ol{\sigma}_0\Id\; .
\end{equation}
With $\rho_0=3\ol{\sigma}_0^2$ and $\xi_0=2\ol{\sigma}_0^3$ it is
parameterized by
\begin{equation}
  \label{ExpansionPotential}
  \begin{array}{rcl}
    \ds{V} &=&
    \ds{\ol{m}_g^2\left(\rho-\rho_0\right)
      -\hal\ol{\nu}\left[\xi-\xi_0-\ol{\si}_0(\rho-\rho_0)\right]+
      \hal\ol{\lambda}_1\left(\rho-\rho_0\right)^2+\hal\ol{\lambda}_2\tau_2+
      \hal\ol{\lambda}_3\tau_3 }\nnn
    &+& \ds{
      \hal\ol{\beta}_1\left(\rho-\rho_0\right)\left(\xi-\xi_0\right)+
      \hal\ol{\beta}_2\left(\rho-\rho_0\right)\tau_2+
      \hal\ol{\beta}_3\left(\xi-\xi_0\right)\tau_2+
      \hal\ol{\beta}_4\left(\xi-\xi_0\right)^2+\ldots}\; .
  \end{array}
\end{equation}
The mass term $\ol{m}_g^2$ vanishes for zero quark masses
\begin{equation}
  \label{TTT03}
  \ol{m}_g^2=\frac{1}{6\ol{\sigma}_0}\Tr\jmath\; .
\end{equation}
Beyond the minimal kinetic term $\sim
Z_\Phi\Tr\prl_\mu\Phi^\dagger\prl^\mu\Phi$ the effective action
contains more complicated terms involving two derivatives of $\Phi$
multiplied by couplings $X_\Phi^-$, $U_\Phi$,
$\tilde{V}_\Phi$, etc.~\cite{JW96-2}. They play an important role for the
phenomenological analysis and will be specified in detail in the next
section. As a consequence, the wave function renormalization constants
for the different $SU_V(3)$ multiplets contained in $\Phi$ are
different, as for instance $Z_m$ and $Z_p$ for the pseudoscalar octet
and singlet, respectively,
\begin{equation}
 \label{A3}
  \begin{array}{rcl}
    \ds{Z_m} &=& \ds{Z_\Phi+X_\Phi^-\ol{\si}_0^2+U_\Phi\ol{\si}_0}\nnn
    \ds{Z_p} &=& \ds{Z_\Phi+X_\Phi^-\ol{\si}_0^2+
      6\tilde{V}_\Phi\ol{\si}_0^4-2U_\Phi\ol{\si}_0}\; .
  \end{array}
\end{equation}
Because of the axial anomaly reflected by the couplings $\ol{\nu}$,
$\ol{\beta}_1$, $\ol{\beta}_3$ and $\ol{\beta}_4$ the pseudoscalar
singlet acquires a mass term even for vanishing current quark masses
\begin{equation}
 \label{L02}
 \begin{array}{rcl}
   \ds{m_p^2} &=& \ds{
     \left(\frac{3}{2}\ol{\nu}\,\ol{\si}_0+\ol{m}_g^2\right)Z_p^{-1}}\; .
 \end{array}
\end{equation}

For a phenomenological determination of the couplings $\ol{\nu}$,
$\ol{\lambda}_2$, etc. (or the related renormalized parameters
$\sigma_0=Z_m^{-1/2}\ol{\sigma}_0$, $\nu=Z_m^{-3/2}(\ol{\nu}+\ldots)$,
$\lambda_2=Z_m^{-2}\ol{\lambda}_2+\ldots$) we refer the reader to
ref.~\cite{JW96-2}. We only recall here that $\ol{\nu}$ is essentially
determined from the $\eta^\prime$--mass, $\ol{\sigma}_0$ and
$\ol{\lambda}_2$ are related to $f_\pi$ and $f_K-f_\pi$ whereas the
non--minimal kinetic couplings are fixed by the decays
$\eta\ra2\gamma$ and $\eta^\prime\ra2\gamma$. Information from the
scalar sector is not needed for a determination of these parameters.
It will turn out that the couplings $\tilde{L}_5$ and $\tilde{L}_8$
are generated by the exchange of $0^{++}$ octet fields and one
therefore expects that the average scalar octet mass
$m_h^2=(2M_{K_o^*}^2+M_{a_o}^2)/3$ plays a decisive role. It will turn
out, however, that the combination $\frac{Z_h}{Z_m}m_h^2$ which
appears in these couplings (with $Z_h$ the wave function
renormalization constant of the scalar octet) only involves parameters
which can be determined from the pseudoscalar sector alone (at least
to the required order in the quark mass expansion):
\begin{equation}
  \label{TTT04}
  \frac{Z_h}{Z_m}m_h^2\simeq\frac{2}{3}
  \frac{Z_p}{Z_m}M_{\eta^\prime}^2+
  3\lambda_2\sigma_0^2\; .
\end{equation}
The couplings $\tilde{L}_4$ and $\tilde{L}_6$ need, in addition,
information on the octet--singlet mass splitting in the scalar sector
which has not yet been determined phenomenologically. The coupling
$\tilde{L}_7=-\sigma_0/(18\nu)$ is essentially fixed by $f_\pi$ and
the $\eta^\prime$ mass. Not much new information is obtained for the
higher derivative couplings $\tilde{L}_1$, $\tilde{L}_2$ and
$\tilde{L}_3$. They are dominated by corresponding higher derivative
terms in the linear model which are, in turn, mainly generated by the
exchange of vector mesons.

We finally want to mention an earlier analysis \cite{EGPR89-1} of the
hypothesis that the couplings $L_i$ are dominated by the exchange of
scalar and vector mesons. Our findings are in overall agreement with
this work. Beyond \cite{EGPR89-1} we determine here the relevant
couplings and wave function renormalization for the scalar octet. This
permits a rather precise quantitative estimate of the low energy
couplings $\tilde{L}_4$, $\tilde{L}_5$, $\tilde{L}_6$, $\tilde{L}_7$
and $\tilde{L}_8$ within our framework.  Possible effects of higher
states and loops are already included in the effective action for the
linear meson model. Within the above assumptions (i), (ii), (iii)
there are no further corrections to these couplings. The low energy
constants $L_i$ have also been estimated within extended
Nambu--Jona-Lasinio models \cite{HK94-1}.

\sect{Nonlinear versus linear sigma model}
\label{NonlinearVersusLinearSigmaModel}

Our starting point is the polar decomposition of the general complex
$3\times3$ matrix $\Phi$ in terms of a hermitian matrix $S$ and a
unitary matrix $U$
\begin{equation}
  \label{A4}
  \Phi=S U\; ,\;\;\;
  S^\dagger=S\; ,\;\;\;
  U^\dagger U={\bf 1}\; .
\end{equation}
Here $U$ contains the nine $0^{-+}$ pseudoscalars in a nonlinear
representation
\begin{equation}
  \label{A5}
  U=\exp\left\{-\frac{i}{3}\vth\right\}\tilde{U}\; ,\;\;\;
  \det\tilde{U}=1\; ,\;\;\;
  \tilde{U}=\exp\left\{\frac{i\lambda_z\Pi_z}{f}\right\}
\end{equation}
with $\lambda_z$ the Gell--Mann matrices in a normalization
$\tr\lambda_z\lambda_y=2\delta_{zy}$ and $f\equiv(2f_K+f_\pi)/3$ the
average pseudoscalar decay constant. In case of chiral symmetry
breaking the matrix $U$ describes the nine Goldstone bosons which
would be massless in the absence of quark masses and the chiral
anomaly. It therefore transforms under $U_L(3)\times U_R(3)$ chiral
flavor transformations according to
\begin{equation}
  \label{A6}
  U\longrightarrow
  \Uc_R U \Uc_L^\dagger\; .
\end{equation}
With respect to these transformations $\Phi$ belongs to a linear
$(\ol{\bf{3}},\bf{3})$ representation and this determines the
transformation properties of $S=\Phi U^\dagger$ as
\begin{equation}
  \label{A7}
  \Phi\longrightarrow
  \Uc_R\Phi\Uc_L^\dagger\; ,\;\;\;
  S\longrightarrow\Uc_R S\Uc_R^\dagger\; .
\end{equation}
Thus $S$ is neutral with respect to $SU_L(3)$ and the Abelian axial
$U_A(1)$ symmetry. It decomposes with respect to $SU_R(3)$ into an
octet (the traceless part) and a singlet ($\sim\tr S$). The nine real
degrees of freedom contained in $S$ describe in a nonlinear
representation the nine $0^{++}$ scalar fields contained in $\Phi$.
With respect to the discrete symmetries $\Cc$ and $\Pc$ the fields
transform according to
\begin{equation}
  \label{A8}
  \begin{array}{lcccccl}
    \ds{\Pc} &:& \ds{
      \Phi\ra\Phi^\dagger} &,& \ds{
      U\ra U^\dagger} &,& \ds{
      S\ra U^\dagger S U}\\[2mm]
    \ds{\Cc} &:& \ds{
      \Phi\ra\Phi^T} &,& \ds{
      U\ra U^T} &,& \ds{
      S\ra U^T S^T U^*}\; .
  \end{array}
\end{equation}
It is easy to visualize the group theoretical properties underlying
the ansatz (\ref{A4}) by observing that an arbitrary complex matrix
$\Phi$ can be brought into a diagonal and real form $\hat{\Phi}={\rm
  diag}(\hat{\Phi}_u,\hat{\Phi}_d,\hat{\Phi}_s)$ by suitable
$U_L(3)\times U_R(3)$ transformations
$\Phi=\hat{\Uc}_R\hat{\Phi}\hat{\Uc}_L^\dagger$. With
$U=\hat{\Uc}_R\hat{\Uc}_L^\dagger$ we can write
$\Phi=\hat{\Uc}_R\hat{\Phi}\hat{\Uc}_R^\dagger U$ and associate
$S=\hat{\Uc}_R\hat{\Phi}\hat{\Uc}_R^\dagger=S^\dagger$. We finally
observe that the decomposition (\ref{A4}) is not unique. By a
redefinition $S\ra SV^{-1}$, $U\ra V U$ the properties (\ref{A4})
remain unchanged if $V$ is unitary and obeys $V S V=S$. The possible
solutions for $V$ depend on $S$. For instance, $S=1$ requires $V^2=1$
whereas $S=0$ is trivially compatible with all $V$. For generic $S$
there is only a discrete number of solutions $V$. We conclude that for
given $S$ we should define $U$ only modulo $V$. We can use this
freedom to require that for $\Phi=\Phi^\dagger$ one has $U=1$. In
particular, in the presence of a real diagonal source for $\Phi$
(corresponding to a real diagonal quark mass matrix) the expectation
value of $\Phi$ is also real and diagonal such that
$\VEV{\Phi}=\VEV{S}$, $\VEV{U}=1$.

We can now express the different pieces in the chirally invariant
effective Lagrangian for $\Phi$ in terms of $U$ and $S$. The four
independent invariants (\ref{TTT01}) on which the potential $V$,
(\ref{ExpansionPotential}), depends are given by
\begin{equation}
  \label{A9}
  \begin{array}{rclcrcl}
    \ds{\rho} &=& \ds{
      \Tr S^2} &,&
    \ds{\xi} &=& \ds{
      2\det S\cos\vth
      }\nnn
    \ds{\tau_2} &=& \ds{
      \frac{3}{2}\Tr\left(
      S^2-\frac{1}{3}\Tr S^2\right)^2 } &,&
    \ds{\tau_3} &=& \ds{
      \Tr\left(
      S^2-\frac{1}{3}\Tr S^2\right)^3}\; .
  \end{array}
\end{equation}
The additional invariant $\omega=i(\det\Phi-\det\Phi^\dagger)$ is
$\Cc\Pc$--odd and may therefore only appear quadratically.  Yet,
$\omega^2$ can be expressed in terms of the invariants (\ref{A9}). The
potential depends therefore on $S$ and the pseudoscalar singlet field
described by $\vth$.

For the simplest form of the effective kinetic term one finds
\begin{equation}
  \label{A10}
  \begin{array}{rcl}
    \ds{\Lc_{\rm kin}(Z_\Phi)} &=& \ds{
    Z_\Phi\Tr\left(\prl^\mu\Phi^\dagger\prl_\mu\Phi\right)=
      Z_\Phi\Bigg\{
      \Tr\left(\prl^\mu S\prl_\mu S\right)+
      \Tr\left(S^2\prl^\mu U\prl_\mu U^\dagger\right)}\nnn
    &+& \ds{
      \Tr\left([S,\prl^\mu S]
      U\prl_\mu U^\dagger\right)\Bigg\} }
  \end{array}
\end{equation}
whereas some non--minimal kinetic invariants introduced in
\cite{JW96-2} read
\begin{equation}
  \label{A11}
  \begin{array}{rcl}
    \ds{\Lc_{\rm kin}(Y_\Phi)} &=& \ds{
      \frac{1}{4}Y_\Phi\prl_\mu\rho\prl^\mu\rho=
      Y_\Phi\Tr\left( S\prl_\mu S\right)
      \Tr\left( S\prl^\mu S\right)
      }\nnn
    \ds{\Lc_{\rm kin}(V_\Phi)} &=& \ds{
      \frac{1}{2}V_\Phi\prl^\mu\xi\prl_\mu\xi=
      2V_\Phi\Bigg\{
      \cos^2\vth\prl^\mu\det S\prl_\mu\det S}\nnn
    &-& \ds{
      2\det S\cos\vth\sin\vth\prl^\mu\det S\prl_\mu\vth+
      \left(\det S\right)^2\sin^2\vth\prl^\mu\vth\prl_\mu\vth
        \Bigg\} }\nnn
    \ds{\Lc_{\rm kin}(\tilde{V}_\Phi)} &=& \ds{
        \frac{1}{2}\tilde{V}_\Phi
        \prl^\mu\omega\prl_\mu\omega =
        2\tilde{V}_\Phi\Bigg\{
        \sin^2\vth\prl^\mu\det S\prl_\mu\det S}\nnn
    &+& \ds{
        2\det S\cos\vth\sin\vth\prl^\mu\det S\prl_\mu\vth+
        \left(\det S\right)^2\cos^2\vth
          \prl^\mu\vth\prl_\mu\vth\Bigg\} }\nnn
    \ds{\Lc_{\rm kin}(X_\Phi^\pm)} &=& \ds{
          -\frac{1}{8}X_\Phi^\pm\Big\{\Tr\left(
          \Phi^\dagger\prl_\mu\Phi\pm\prl_\mu\Phi^\dagger\Phi\right)
          \left(\Phi^\dagger\prl^\mu\Phi\pm\prl^\mu\Phi^\dagger\Phi\right)
            }\nnn
    && \ds{\hspace{11.5mm}+
      \Tr\left(\Phi\prl_\mu\Phi^\dagger\pm\prl_\mu\Phi\Phi^\dagger\right)
      \left(\Phi\prl^\mu\Phi^\dagger\pm\prl^\mu\Phi\Phi^\dagger\right)
        \Big\} }\nnn
    &=& \ds{-\frac{1}{4}X_\Phi^\pm\Bigg\{
      2\Tr\left( S\prl^\mu S S\prl_\mu S\pm
      S^2\prl^\mu S\prl_\mu S\right)}\nnn
    && \ds{\hspace{1.4cm}
      +\Tr\left\{\left(\left(2\mp1\right)
      [S\prl^\mu S S,S]\mp [S^3,\prl^\mu S]\right)
      \prl_\mu U U^\dagger\right\} }\nnn
    && \ds{\hspace{1.4cm}
      -\Tr\left\{\left(2\mp1\right) S^2 U\prl^\mu U^\dagger
      S^2 \prl_\mu U U^\dagger\mp
      S^4 \prl^\mu U\prl_\mu U^\dagger
    \right\} \Bigg\} }\; .
  \end{array}
\end{equation}
We also include an ``anomalous'' $U_A(1)$ violating kinetic term
involving $\epsilon$--tensors
\begin{equation}
  \label{A12}
  \begin{array}{rcl}
    \ds{\Lc_{\rm kin}(U_\Phi)} &=& \ds{
      \frac{1}{2}U_\Phi
      \eps^{a_1a_2a_3}\eps^{b_1b_2b_3}
      \left(\Phi_{a_1b_1}\prl^\mu\Phi_{a_2b_2}\prl_\mu\Phi_{a_3b_3}+
        \Phi^\dagger_{a_1b_1}\prl^\mu\Phi^\dagger_{a_2b_2}\prl_\mu
        \Phi^\dagger_{a_3b_3}
      \right)}\nnn
    &=& \ds{
      U_\Phi\cos\vth\Bigg\{
      \Tr S\Tr\prl_\mu S\Tr\prl^\mu S+
      2\Tr\left( S\prl_\mu S\prl^\mu S\right)-
      \Tr S\Tr\left( \prl_\mu S\prl^\mu S\right)
      }\nnn
    &-& \ds{
      2\Tr\prl_\mu S\Tr\left( S\prl^\mu S\right)+
      \Tr\left(\prl_\mu S\left[
      S^2,\prl^\mu U U^\dagger\right]\right)-
      \Tr S\Tr\left(
      \prl_\mu S\left[ S,\prl^\mu U U^\dagger\right]\right)}\nnn
    &+& \ds{
      \det S
      \left(\Tr\left(\prl_\mu U\prl^\mu U^\dagger\right)-
        \prl_\mu\vth\prl^\mu\vth\right)\Bigg\} }\nnn
    &-& \ds{
      i U_\Phi\sin\vth\Bigg\{
      2\Tr S\Tr\prl_\mu S\Tr\left( S\prl^\mu U U^\dagger\right)-
      \Tr S\Tr\left(\prl_\mu S 
      \left\{ S,\prl^\mu U U^\dagger\right\}\right)
      }\nnn
    &-& \ds{
      2\Tr\prl_\mu S\Tr\left( S^2
      \prl^\mu U U^\dagger\right) -
      2\Tr\left(\prl_\mu U U^\dagger S\right)
      \Tr\left( S\prl^\mu S\right)}\nnn
    &+& \ds{
      \Tr\left(\prl_\mu S\left( S
      \left\{ S,\prl^\mu U U^\dagger\right\}+
        \left\{ S,\prl^\mu U U^\dagger\right\} S\right)\right)
      \Bigg\}
      }
  \end{array}
\end{equation}
where we used the relation
\begin{equation}
  \label{VVV01}
  \begin{array}{rcl}
    \ds{\epsilon^{a_1 a_2 a_3}\epsilon^{b_1 b_2 b_3}
      A_{a_1 b_1}B_{a_2 b_2}C_{a_3 b_3}} &=& \ds{
      \Tr A\Tr B\Tr C+
      \Tr\left( A\left\{ B,C\right\}\right)}\nnn
    &-& \ds{
      \Tr A\Tr\left( B C\right)-
      \Tr B\Tr\left( A C\right)-
      \Tr C\Tr\left( A B\right)}\; .
  \end{array}
\end{equation}
In addition, there are terms involving four and more derivatives of
$\Phi$.

The effective action of the nonlinear sigma model obtains if we
restrict the degrees of freedom of the linear model to the
pseudoscalars in the nonlinear realization $U$.  A straightforward
approach inserts (\ref{A4}), (\ref{A5}) into the potential $V$ and the
kinetic terms of the linear meson model. Subsequently one expresses
$S$ as a functional of $U$ by means of the solution of its equation of
motion.  This is, of course a rather difficult task. On the other
hand, we are only interested in the effective action for $U$ or
$\tilde{U}$ to low orders in an expansion in powers of derivatives and
$M_q$. This reduces the complexity of this program considerably.

\sect{Lowest order chiral perturbation theory}
\label{LowestOrderNonlinearSigmaModel}

We will start by considering first the lowest order of the quark mass
expansion of the effective action for the linear meson model. This
coincides with lowest order chiral perturbation theory.  In this
approximation $\Phi$ can be written as
\begin{equation}
 \Phi=\vph_{00}U\; ,\;\;\;\;
 {\rm i.e.}\;\;\; S=\vph_{00}\Id\; .
 \label{PhiPhi}
\end{equation}
Here the positive real constant $\vph_{00}$ should be associated with
the expectation value of $\Phi$ in the limit of vanishing quark
masses. Only in this limit $\vph_{00}$ coincides with $\ol{\si}_0$ and
$\VEV{\Phi}-\vph_{00}=0$, whereas for $m_q\neq0$ there are corrections
involving the trace of the quark mass matrix.

The computation of the parameters of chiral perturbation theory from
those of the linear meson model proceeds in two steps: First one
inserts eqs.~(\ref{PhiPhi}) with (\ref{A5}) into the kinetic terms
(\ref{A10})--(\ref{A12}), the potential (\ref{ExpansionPotential}) and
the source term (\ref{SourceTerm}) linear in the quark masses.
Second, one establishes the relation between $\vph_{00}$ and
$\ol{\si}_0$.  This yields an effective action for $\tilde{U}$ and
$\vth$ in the limit of small quark masses.  Within our approach the
parameters in this effective action correspond to an expansion in
momenta around a typical momentum in the vicinity of the pole for the
pseudoscalar mesons, i.e. $q_0^2=-(2M_{K^\pm}^2+M_{\pi^\pm}^2)/3$.

Let us start with the kinetic terms which yield
\begin{equation}
 \label{K11}
  \begin{array}{rcl}
    \ds{\Lc_{\rm kin}^{(0)}} &=& \ds{
      \left( Z_\Phi+U_\Phi\vph_{00}\cos\vth+
        X_\Phi^-\vph_{00}^2\right)
        \vph_{00}^2 \Tr\prl^\mu U^\dagger \prl_\mu U}\nnn
    &+& \ds{
      \vph_{00}^3 \prl^\mu\vth\prl_\mu\vth
      \left( 2\vph_{00}^3 V_\Phi\sin^2\vth+
        2\vph_{00}^3 \tilde{V}_\Phi\cos^2\vth-
      U_\Phi\cos\vth\right) } \; .
  \end{array}
\end{equation}
With the identities
\begin{equation}\begin{array}{rcl}
 \ds{\Tr\left(\tilde{U}^\dagger\prl_\mu\tilde{U}\right)} &=& 0\;
 ,\;\;\;
 \Tr\left(\prl_\mu U U^\dagger\right)=-i\prl_\mu\vth\nnn
 \ds{\Tr\prl^\mu U^\dagger\prl_\mu U} &=& \ds{
 \Tr\prl^\mu\tilde{U}^\dagger\prl_\mu\tilde{U}+
 \frac{1}{3}\prl^\mu\vth\prl_\mu\vth}
\end{array}\end{equation}
we find the singlet kinetic term from (\ref{K11}) for
\begin{equation}
  \begin{array}{rcl}
    \ds{\Lc_{\rm kin}^{(0)}(\vth)} &=& \ds{
      \frac{1}{3}\vph_{00}^2\prl^\mu\vth\prl_\mu\vth\Bigg\{
      Z_\Phi+X_\Phi^-\vph_{00}^2}\nnn
    &-& \ds{
      2U_\Phi\vph_{00}\cos\vth+6\vph_{00}^4
      \left(V_\Phi\sin^2\vth+\tilde{V}_\Phi\cos^2\vth\right)\Bigg\} }\; .
  \end{array}
\end{equation}
We note that at $\vth=0$ the kinetic term is positive provided
$Z_p^{(0)}=Z_\Phi+X_\Phi^-\vph_{00}^2-2U_\Phi\vph_{00}+
6\tilde{V}_\Phi\vph_{00}^4>0$.  We recognize the wave function
renormalization $Z_p$, (\ref{A3}), up to the difference between
$\vph_{00}$ and $\ol{\sigma}_0$. Similarly, the combination
$Z_m^{(0)}=Z_\Phi+U_\Phi\vph_{00}+X_\Phi^-\vph_{00}^2$ in front of the
octet kinetic term $\sim\Tr\prl^\mu\tilde{U}^\dagger\prl_\mu\tilde{U}$
in (\ref{K11}) is the lowest order approximation of $Z_m$.
(\ref{A3}). Instead of $\vth$ we may use a field $p$ with standard
normalization of the kinetic term
\begin{equation}
  \label{BBB55}
  \frac{1}{2}\prl^\mu p\prl_\mu p=
  \frac{1}{3}\vph_{00}^2 Z_p
  \prl^\mu\vth\prl_\mu\vth\; .
\end{equation}

With (\ref{PhiPhi}) the effective potential $V$
(\ref{ExpansionPotential}) is independent of $\tilde{U}$ and
contributes only a potential for $\vth$
\begin{equation}
  \label{ThetaPotential}
  \begin{array}{rcl}
    \ds{V(\vth)} &=& \ds{
      -\hal\left[\ol{\nu}+
      3\ol{\beta}_1(\ol{\si}_0^2-\vph_{00}^2)+
      2\ol{\beta}_4\ol{\si}_0^3\right]\xi(\vth)+
      \hal\ol{\beta}_4\xi(\vth)^2}\nnn
    \ds{\xi(\vth)} &=& \ds{
      2\vph_{00}^3\cos\vth}\; .
  \end{array}
\end{equation}
Expanding around the minimum at $\vth=0$ this induces a mass term for
the pseudoscalar singlet
\begin{equation}
  \label{EtapMassTerm}
  \begin{array}{rcl}
    \ds{\frac{1}{2}\ol{M}_\vth^2\vth^2} &=& \ds{
      \hal\vph_{00}^3\left[\ol{\nu}+
      3\ol{\beta}_1(\ol{\si}_0^2-\vph_{00}^2)+
      2\ol{\beta}_4(\ol{\si}_0^3-2\vph_{00}^3)
    \right]\vth^2}\nnn
  &=& \ds{
    \hal p^2Z_p^{-1}
    \left[\frac{3}{2}\ol{\nu}\vph_{00}+
      \frac{9}{2}\ol{\beta}_1\vph_{00}
      (\ol{\si}_0^2-\vph_{00}^2)+
      3\ol{\beta}_4\vph_{00}
      (\ol{\si}_0^3-2\vph_{00}^3)
    \right]=\frac{1}{2}m_p^{(0)2} p^2}\; .
  \end{array}
\end{equation}
Here $m_p^{(0)}$ can be identified to lowest order with $m_p$,
(\ref{L02}) or the mass of the $\eta^\prime$ meson.  For vanishing
quark masses we therefore have a massless octet of Goldstone bosons
$\tilde{U}$ and a massive pseudoscalar singlet $p$.  The source term
(\ref{SourceTerm})
\begin{equation}
  \label{STLowestOrder}
  \Lc_\jmath^{(0)}=-\hal\vph_{00}\Tr
  \left( \jmath^\dagger U+U^\dagger \jmath\right)=
    -\hal a_q\vph_{00}\Tr M_q
    \left( U+U^\dagger\right)
\end{equation}
generates non--vanishing masses for the pseudoscalar octet.

It is now straightforward to identify the parameters of the nonlinear
sigma model to lowest order in the quark mass expansion.  We use here
the notation of \cite{GL82-1} adapted to Euclidean space--time
\begin{equation}
  \label{TTT05}
    \ds{\Lc^{(0)}_{\rm\chi PT}} = \ds{
      \frac{\tilde{F}_0^2}{4}\left\{\Tr\left(
      \prl_\mu U\prl^\mu U^\dagger\right)-
      2\tilde{B}_0\Tr M_q\left( U+U^\dagger\right)
      \right\}+
      \frac{1}{12}\tilde{H}_0\prl^\mu\vth\prl_\mu\vth }\; .
\end{equation}
We find
\begin{equation}
  \label{GLParameters}
  \begin{array}{rcl}
    \ds{\tilde{F}_0^2} &=& \ds{
      4\vph_{00}^2
      \left( Z_\Phi+U_\Phi\vph_{00}+ X_\Phi^-\vph_{00}^2\right)}\nnn
    \ds{\tilde{B}_0} &=& \ds{
      \frac{1}{4}\frac{a_q}
      {\left( Z_\Phi+U_\Phi\vph_{00}+ X_\Phi^-\vph_{00}^2\right)
        \vph_{00}}}\nnn 
    \ds{\tilde{H}_0} &=& \ds{ 
      -12U_\Phi\vph_{00}^3+
      24\tilde{V}_\Phi\vph_{00}^6
      }
  \end{array}
\end{equation}
and $f$ in (\ref{A5}) is given to this order by
\begin{equation}
  \label{BBB80}
  f=2Z_m^{1/2}\vph_{00}\; .
\end{equation}
The constants $\tilde{F}_0^2$ and $\tilde{H}_0$ can be computed from
the parameters of the linear meson model once the relation between
$\vph_{00}$ and $\ol{\si}_0$ is established.  To lowest order in the
quark mass expansion we can use
$\vph_{00}=\ol{\sigma}_0=Z_m^{-1/2}(2f_K+f_\pi)/6$. For quantitative
estimates in next to leading order we also need to include the
difference between $\vph_{00}$ and $\ol{\sigma}_0$.  The value of
$\vph_{00}$ corresponds to the minimum of the effective potential
(\ref{ExpansionPotential}) without the source term. We only need to
evaluate $V$ for diagonal fields $\Phi=\ol{\si}_0+\Dt\si$
\begin{equation}
  \label{L03}
  \begin{array}{rcl}
  \ds{V(\Dt\si)} &=& \ds{
  6\ol{m}_g^2\ol{\si}_0\Dt\si+
  \left(3\ol{m}_g^2-\frac{3}{2}\ol{\nu}\,\ol{\si}_0+
  18\ol{\lambda}_1\ol{\si}_0^2+
  18\ol{\beta}_1\ol{\si}_0^3+
  18\ol{\beta}_4\ol{\si}_0^4
  \right)\Dt\si^2}\nnn
  &+& \ds{
  \left(-\ol{\nu}+18\ol{\lambda}_1\ol{\si}_0+
  27\ol{\beta}_1\ol{\si}_0^2+
  36\ol{\beta}_4\ol{\si}_0^3
  \right)\Dt\si^3 }\nnn
  &+& \ds{
  \left(\frac{9}{2}\ol{\lambda}_1+
  15\ol{\beta}_1\ol{\si}_0+
  30\ol{\beta}_4\ol{\si}_0^2
  \right)\Dt\si^4+
  \left(3\ol{\beta}_1+
  12\ol{\beta}_4\ol{\si}_0
  \right)\Dt\si^5+
  2\ol{\beta}_4\Dt\si^6}\; .
  \end{array}
\end{equation}
Including corrections to quadratic order in $\Dt\si$ this yields
\begin{equation}
  \label{L04}
  \begin{array}{rcl}
    \ds{\vph_{00}} &=& \ds{
      \ol{\si}_0-\ol{\si}_0\frac{\ol{m}_g^2}{\ol{m}_s^2}+
      \hal\frac{\ol{m}_g^4\ol{\si}_0^2}{\ol{m}_s^6}
      \left(\ol{\nu}-
        18\ol{\lambda}_1\ol{\si}_0-
        27\ol{\beta}_1\ol{\si}_0^2-
        36\ol{\beta}_4\ol{\si}_0^3
      \right)}
  \end{array}
\end{equation}
where
\begin{equation}
  \label{TTT08}
  \begin{array}{rcl}
    \ds{\ol{m}_s^2} &=& \ds{
      \ol{m}_g^2-
        \hal\ol{\nu}\,\ol{\si}_0+
        6\ol{\lambda}_1\ol{\si}_0^2+
        6\ol{\beta}_1\ol{\si}_0^3+
        6\ol{\beta}_4\ol{\si}_0^4
      }
  \end{array}
\end{equation}
equals the scalar singlet mass term $m_s^2$ up to a wave function
renormalization constant ($m_s^2=\ol{m}_s^2/Z_s$).  Inserting
(\ref{L04}) into (\ref{GLParameters}) one can now extract the
difference between $\tilde{F}_0$ and $f=(2f_K+f_\pi)/3$ to quadratic
order in the quark masses.  The dominant contribution (neglecting
corrections $\sim U_\Phi$ and $X_\Phi^-$) is
$\tilde{F}_0/f\simeq\vph_{00}/\ol{\sigma}_0\simeq1-\ol{m}_g^2/\ol{m}_s^2$.
We remind the reader that according to our conventions for the
definition of $Z_\Phi$, $X_\Phi^-$, etc.~the parameter $\tilde{F}_0^2$
corresponds here to a normalization at $q_0^2$. The value of
$\tilde{B}_0$ is related to the proportionality constant\footnote{Note
  that only the combination $a_qM_q$ is independent of the
  renormalization scale used for the definition of the current quark
  masses.} $a_q$ in (\ref{Current}). It sets the scale for the average
current quark mass to lowest order. An estimate of the current quark
masses and $\tilde{B}_0$ in the linear meson model can be found in
\cite{JW96-3}.

Finally, it is instructive to relate the fields of the nonlinear sigma
model to the linear representations contained in $\Phi$:
\begin{equation}
  \label{BBB56}
  \Phi=\ol{\sigma}_0+\frac{1}{\sqrt{2}}
  \left( i\phi_p+\frac{i}{\sqrt{3}}\chi_p+
    \phi_s+\frac{1}{\sqrt{3}}\chi_s\right)\; .
\end{equation}
To lowest order one may use for $\VEV{\Phi}=\vph_{00}$ the relations
\begin{equation}
  \label{BBB51}
  \begin{array}{rcl}
    \ds{\chi_p} &=& \ds{-\frac{i}{\sqrt{6}}\vph_{00}
      \Tr\left( U-U^\dagger\right) }\nnn
    \ds{\Phi_p} &=& \ds{
      -\frac{i}{\sqrt{2}}\vph_{00}
      \left[ U- U^\dagger-\frac{1}{3}
        \Tr\left( U- U^\dagger\right)\right]}\nnn
    \ds{\chi_s} &=& \ds{
      \frac{1}{\sqrt{6}}\vph_{00}
      \left[\Tr\left( U+ U^\dagger\right)-6\right]}\nnn
    \ds{\Phi_s} &=& \ds{
      \frac{1}{\sqrt{2}}\vph_{00}
      \left[ U+ U^\dagger-\frac{1}{3}
        \Tr\left( U+ U^\dagger\right)\right]}
  \end{array}
\end{equation}
which implies for the kinetic terms
\begin{equation}
  \label{BBB52}
  \begin{array}{rcl}
    \ds{\Tr\prl^\mu\Phi_p\prl_\mu\Phi_p} &=& \ds{
      \hal\vph_{00}^2\Bigg\{
      2\Tr\prl^\mu U^\dagger\prl_\mu U-
      \Tr\prl^\mu U\prl_\mu U}\nnn
    &-&\ds{
      \Tr\prl^\mu U^\dagger\prl_\mu U^\dagger+
      \frac{1}{3}\left[\Tr(\prl_\mu U-\prl_\mu U^\dagger)\right]^2
      \Bigg\} }\nnn
    \ds{\prl^\mu\chi_p\prl_\mu\chi_p} &=& \ds{
      -\frac{1}{6}\vph_{00}^2
      \left[\Tr(\prl_\mu U-\prl_\mu U^\dagger)\right]^2}\nnn
    \ds{\Tr\prl^\mu\Phi_s\prl_\mu\Phi_s} &=& \ds{
      \hal\vph_{00}^2\Bigg\{
      2\Tr\prl^\mu U^\dagger\prl_\mu U+
      \Tr\prl^\mu U\prl_\mu U}\nnn
    &+& \ds{
      \Tr\prl^\mu U^\dagger\prl_\mu U^\dagger-
      \frac{1}{3}\left[\Tr(\prl_\mu U+\prl_\mu U^\dagger)\right]^2
      \Bigg\} }\nnn
    \ds{\prl^\mu\chi_s\prl_\mu\chi_s} &=& \ds{
      \frac{1}{6}\vph_{00}^2
      \left(\Tr(\prl_\mu U+\prl_\mu U^\dagger)\right)^2}\; .
  \end{array}
\end{equation}
On the other hand,
expanding $U$ for small $\Pi_z$ and $\vth$
yields
\begin{equation}
  \label{BBB53}
  \begin{array}{rcl}
    \ds{\vph_{00}\tilde{U}} &=& \ds{
      \vph_{00}+\frac{i}{2}Z_m^{-\hal}\la_z\Pi_z-
      \frac{1}{8}Z_m^{-1}\frac{(\la_z\Pi_z)^2}{\vph_{00}}+\ldots}\nnn
    \ds{\vph_{00}U} &=& \ds{
      \vph_{00}\tilde{U}-\frac{i}{3}\vph_{00}\vth-
      \frac{1}{18}\vph_{00}\vth^2+
      \frac{1}{6}Z_m^{-\hal}\vph_{00}\vth\la_z\Pi_z+\ldots }\; .
  \end{array}
\end{equation}
To linear order in $\Pi$ and $\vth$ one can identify $\Pi$ with
$m=(2Z_m)^{1/2}\phi_p$
\begin{equation}
 \Pi=\la_z\Pi_z=(2Z_m)^\hal\Phi_p=\la_z m_z
\end{equation}
and $\vth$ is proportional to $p=Z_p^{1/2}\chi_p$
\begin{equation}
 \vth=-\frac{3}{\sqrt{6}}\frac{\chi_p}{\vph_{00}}=
 -\frac{3}{\sqrt{6}}Z_p^{-\hal}\frac{p}{\vph_{00}}\; .
\end{equation}

\sect{Next to leading order quark mass expansion}
\label{NextToLeadingOrder}

Going beyond the lowest order in chiral perturbation theory requires a
solution for $S[U]$ in the presence of ``sources'' involving quark
masses, non--vanishing $\vth$ and derivatives of $U$. We will
construct an iterative solution $S[U]$ treating the sources as small
corrections. Inserting this solution into the action will produce a
systematic expansion of the effective action of the nonlinear sigma
model. The lowest order of this expansion is $S=\vph_{00}$ and has
been discussed in the previous section. Next one expands the general
solution for $U\neq1$, namely,
$S=\vph_{00}+S_1+\ldots\equiv\vph_{00}+H+T+\ldots$, with $\Tr H=0$ and
$T=(\Tr S_1)/3$, while keeping only terms linear in $S_1$ and the
sources. To this order one needs the Lagrangian for $H$ and $T$ in the
approximation
\begin{equation}
  \label{BBB01}
  \begin{array}{rcl}
    \ds{\Lc^{(H,T)}} &=& \ds{
      \frac{\ol{M}_H^2}{2}\Tr H^2
      -\Tr A^{(H)}[U]H+
      \frac{\ol{M}_T^2}{2}T^2
      -A^{(T)}[U]T}
  \end{array}
\end{equation}
The mass terms $\ol{M}_H^2$ and $\ol{M}_T^2$ obtain from the second
derivatives of $V$ at $S=\vph_{00}$ (to leading order we may set
$\vph_{00}=\ol{\sigma}_0$ here):
\begin{equation}
  \label{BBB02}
  \begin{array}{rcl}
    \ds{\ol{M}_H^2} &=& \ds{
      2\ol{m}_g^2+2\ol{\nu}\,\ol{\sigma}_0+
      6\ol{\lambda}_2\ol{\sigma}_0^2
      }\nnn
    \ds{\ol{M}_T^2} &=& \ds{
      6\ol{m}_g^2-3\ol{\nu}\,\ol{\sigma}_0+
      36\ol{\lambda}_1\ol{\sigma}_0^2+
      36\ol{\beta}_1\ol{\sigma}_0^3+
      36\ol{\beta}_4\ol{\sigma}_0^4=
      6\ol{m}_s^2
      }\; .
  \end{array}
\end{equation}
The source terms have three different contributions. The first arises
from the quark mass term
\begin{equation}
  \label{BBB03}
  \begin{array}{rcl}
    \ds{A_\jmath^{(H)}} &=& \ds{
      \frac{1}{2}\left( U\jmath^\dagger+\jmath U^\dagger\right)-
      \frac{1}{6}\Tr\left( U\jmath^\dagger+\jmath U^\dagger\right)
      }\nnn
    \ds{A_\jmath^{(T)}} &=& \ds{
      \frac{1}{2}\Tr\left( U\jmath^\dagger+\jmath U^\dagger\right)
      }\; .
  \end{array}
\end{equation}
The second contribution obtains from linearizing the kinetic term in
$H+T$
\begin{equation}
  \label{BBB04}
  \begin{array}{rcl}
    \ds{A_{\rm kin}^{(H)}} &=& \ds{
      -2\vph_{00}\left( Z_\Phi+
      2X_\Phi^-\vph_{00}^2\right)
      \left(\prl_\mu U\prl^\mu U^\dagger-
        \frac{1}{3}\Tr\prl_\mu U\prl^\mu U^\dagger\right)
      }\nnn
    &-& \ds{
      2i U_\Phi\vph_{00}^2\left(
      \prl_\mu\left[\sin\vth U\prl^\mu U^\dagger\right]-
      \frac{i}{3}\prl_\mu
      \left[\sin\vth\prl^\mu\vth\right]
      \right)
      }\nnn
    \ds{A_{\rm kin}^{(T)}} &=& \ds{
      -\vph_{00}\left\{
      2\left( Z_\Phi+2X_\Phi^-\vph_{00}^2\right)+
      3U_\Phi\vph_{00}\cos\vth\right\}
      \Tr\prl_\mu U\prl^\mu U^\dagger
      }\nnn
    &-& \ds{
      \vph_{00}^2\left\{
      12\vph_{00}^3\left(V_\Phi\cos^2\vth+
      \tilde{V}_\Phi\sin^2\vth\right)+
      U_\Phi\cos\vth\right\}
      \prl_\mu\vth\prl^\mu\vth
      }\nnn
    &-& \ds{
      4\vph_{00}^2\left\{3\vph_{00}^3\sin\vth\cos\vth
      \left( V_\Phi-\tilde{V}_\Phi\right)+
      U_\Phi\sin\vth\right\}
        \prl^\mu\prl_\mu\vth
      }\; .
  \end{array}
\end{equation}
Finally, the last term results from the $\vth$--dependence of
$V$ induced by the invariant $\xi$ (with $\vph_{00}$ replaced by
$\ol{\sigma}_0$ here)
\begin{equation}
  \label{BBB05}
  \begin{array}{rcl}
    \ds{A_{\vth}^{(H)}} &=& \ds{0}\nnn
    \ds{A_{\vth}^{(T)}} &=& \ds{
      -3\ol{\nu}\,\ol{\sigma}_0^2
      \left(1-\cos\vth\right)
      }\; .
  \end{array}
\end{equation}
We point out that we have neglected in (\ref{BBB01}) kinetic terms for
$H$ and $T$ as induced by $\Lc_{\rm kin}$. This is consistent with our
approximation since an expansion of the propagator $(\ol{M}^2+Z
q^2)^{-1}$ in powers of the momentum squared $q^2$ produces higher
derivatives.

The solution
\begin{equation}
  \label{BBB06}
  \begin{array}{rcl}
    \ds{H} &=& \ds{
      \ol{M}_H^{-2}A^{(H)}[U]}\nnn
    \ds{T} &=& \ds{
      \ol{M}_T^{-2}A^{(T)}[U]}
  \end{array}
\end{equation}
has now to be reinserted into the effective action. This yields the
next to leading order corrections to the effective Lagrangian of
the nonlinear sigma model
\begin{equation}
  \label{BBB07}
  \Lc_{\chi PT}^{(1)}=
  -\frac{1}{2\ol{M}_H^2}\Tr\left(
  A_\jmath^{(H)}+A_{\rm kin}^{(H)}+A_\vth^{(H)}\right)^2
  -\frac{1}{2\ol{M}_T^2}\left(
  A_\jmath^{(T)}+A_{\rm kin}^{(T)}+A_\vth^{(T)}\right)^2\; .
\end{equation}
We will write the terms which do not involve $\vth$ explicitly (i.e.
beyond the implicit $\vth$--dependence of $U$, $U^\dagger$) in a
notation analogous to that of ref.~\cite{GL82-1} (for Euclidean
space--time)
\begin{equation}
  \label{BBB57}
  \begin{array}{rcl}
    \ds{\Lc_{\rm\chi PT}^{(1)}} &=& \ds{
      -\hat{L}_1\left[\Tr\left(
      \prl_\mu U\prl^\mu U^\dagger\right)\right]^2
      -\hat{L}_2\Tr\left(\prl_\mu U\prl_\nu U^\dagger\right)
      \Tr\left(\prl^\mu U\prl^\nu U^\dagger\right)}\nnn
    &-& \ds{
      \hat{L}_3\Tr\left(\prl_\mu U\prl^\mu U^\dagger\right)^2+
      2 B_0 \hat{L}_4\Tr\left(\prl_\mu U\prl^\mu U^\dagger\right)
      \Tr M_q\left( U^\dagger+U\right)}\nnn
    &+& \ds{
      2B_0 \hat{L}_5\Tr\left(\prl_\mu U\prl^\mu U^\dagger
      \left[ M_q U^\dagger+U M_q\right]\right)-
      4B_0^2 \hat{L}_6\left[
      \Tr M_q\left( U^\dagger+U\right)\right]^2}\nnn
    &-& \ds{
      4B_0^2 \hat{L}_7\left[
      \Tr M_q\left( U^\dagger-U\right)\right]^2-
      4B_0^2 \hat{L}_8\Tr\left(
      M_q U^\dagger M_q U^\dagger+
      M_q U M_q U\right)}\; .
  \end{array}
\end{equation}
The coefficients $\hat{L}_i$ can be read off from (\ref{BBB07}) by
setting $\vth=0$.  The next to leading order quark mass corrections in
the potential for $U$ arise from the terms $\sim A_\jmath^2$
\begin{equation}
  \label{BBB08}
  \Lc_m^{(1)}=-\frac{1}{8\ol{M}_H^2}
  \left(\Tr\jmath^\dagger U\jmath^\dagger U+
    \Tr U^\dagger\jmath U^\dagger\jmath\right)-
  \frac{1}{8}\left(\frac{1}{\ol{M}_T^2}-\frac{1}{3\ol{M}_H^2}\right)
  \left[\Tr\left(\jmath^\dagger U+U^\dagger\jmath \right)\right]^2\; .
\end{equation}
Comparing with (\ref{BBB57}) we infer the constants
\begin{equation}
  \label{BBB09}
  \begin{array}{rcl}
    \ds{\hat{L}_6} &=& \ds{
      \frac{1}{2}
      \left(Z_\Phi+U_\Phi\vph_{00}+X_\Phi^-\vph_{00}^2\right)^2
      \left(\frac{\vph_{00}^2}{\ol{M}_T^2}-
        \frac{\vph_{00}^2}{3\ol{M}_H^2}\right)
      }\nnn
    \ds{\hat{L}_7} &=& \ds{0}\nnn
    \ds{\hat{L}_8} &=& \ds{
      \frac{1}{2}
      \left(Z_\Phi+U_\Phi\vph_{00}+X_\Phi^-\vph_{00}^2\right)^2
      \frac{\vph_{00}^2}{\ol{M}_H^2}
      }\; .
  \end{array}
\end{equation}
Quark mass corrections to the kinetic terms are induced by $A_\jmath
A_{\rm kin}$ and read
\begin{equation}
  \label{BBB10}
  \begin{array}{rcl}
    \ds{\Lc_{\rm kin,m}^{(1)}} &=& \ds{
      \left(Z_\Phi+2X_\Phi^-\vph_{00}^2\right)
        \frac{\vph_{00}}{\ol{M}_H^2}\Tr\left[\left(
      U\jmath^\dagger+\jmath U^\dagger\right)
      \prl_\mu U\prl^\mu U^\dagger\right]}\nnn
    &+& \ds{
      \left[\left(Z_\Phi+\frac{3}{2}U_\Phi\vph_{00}+
        2X_\Phi^-\vph_{00}^2\right)
        \frac{\vph_{00}}{\ol{M}_T^2}-
        \frac{1}{3}\left(Z_\Phi+
        2X_\Phi^-\vph_{00}^2\right)
          \frac{\vph_{00}}{\ol{M}_H^2}\right]
      }\nnn
    &\times& \ds{
          \Tr\left(
      U\jmath^\dagger+\jmath U^\dagger\right)
      \Tr\prl_\mu U\prl^\mu U^\dagger}\; .
  \end{array}
\end{equation}
From there we find the coefficients
\begin{equation}
  \label{BBB59}
  \begin{array}{rcl}
    \ds{\hat{L}_4} &=& \ds{2
      \left(Z_\Phi+U_\Phi\vph_{00}+X_\Phi^-\vph_{00}^2\right)
        }\nnn
    &\times& \ds{
      \left[\left(Z_\Phi+\frac{3}{2}U_\Phi\vph_{00}+
        2X_\Phi^-\vph_{00}^2\right)
        \frac{\vph_{00}^2}{\ol{M}_T^2}-
        \frac{1}{3}\left(Z_\Phi+2X_\Phi^-\vph_{00}^2\right)
        \frac{\vph_{00}^2}{\ol{M}_H^2}\right]
      }\nnn
    \ds{\hat{L}_5} &=& \ds{2
      \left(Z_\Phi+U_\Phi\vph_{00}+X_\Phi^-\vph_{00}^2\right)
      \left(Z_\Phi+2X_\Phi^-\vph_{00}^2\right)
      \frac{\vph_{00}^2}{\ol{M}_H^2}
      }\; .
  \end{array}
\end{equation}

Finally, we turn to the terms involving four derivatives of $U$.  In
the linear sigma model they obtain also contributions from invariants
involving four derivatives of $\Phi$. For instance, a term of the form
\begin{equation}
  \label{BBB60}
  \Lc=\ol{H}_\Phi\Tr
  \left(\prl^\mu\Phi^\dagger\prl_\mu\Phi\right)^2
\end{equation}
yields
\begin{equation}
  \label{BBB61}
  \Lc_{\rm kin,4}^{(0)}=\ol{H}_\Phi\vph_{00}^4\Tr
  \left(\prl^\mu U^\dagger\prl_\mu U\right)^2\; .
\end{equation}
Since we are considering a linear meson model without explicit vector
and axial--vector fields it is plausible that the dominant part of
$\ol{H}_\Phi$ and also $\hat{L}_3$ (as well as $\hat{L}_1$ and
$\hat{L}_2$) is generated by the exchange of vector mesons
\cite{EGPR89-1}. Additional contributions are produced by scalar
exchange $\sim A_{\rm kin}^2$:
\begin{equation}
  \label{BBB62}
  \begin{array}{rcl}
    \ds{\Lc_{\rm kin,4}^{(1)}} &=& \ds{
      -2\left(Z_\Phi+2X_\Phi^-\vph_{00}^2\right)^2
        \frac{\vph_{00}^2}{\ol{M}_H^2}
      \left\{\Tr\left(\prl^\mu U\prl_\mu U^\dagger\right)^2-
        \frac{1}{3}\left[\Tr
      \left(\prl_\mu U\prl^\mu U^\dagger\right)\right]^2\right\}
      }\nnn
    &-& \ds{
      2\left(Z_\Phi+\frac{3}{2}U_\Phi\vph_{00}+
        2X_\Phi^-\vph_{00}^2\right)^2
        \frac{\vph_{00}^2}{\ol{M}_T^2}
          \left[\Tr
      \left(\prl_\mu U\prl^\mu U^\dagger\right)\right]^2
      }\; .
  \end{array}
\end{equation}
leading to
$\hat{L}_i$ are
\begin{equation}
  \label{BBB63}
  \begin{array}{rcl}
    \ds{ \hat{L}_1^{(H,T)}} &=& \ds{2
      \left(Z_\Phi+\frac{3}{2}U_\Phi\vph_{00}+2X_\Phi^-\vph_{00}^2\right)^2
      \frac{\vph_{00}^2}{\ol{M}_T^2}-
      \frac{2}{3}\left(Z_\Phi+2X_\Phi^-\vph_{00}^2\right)^2
      \frac{\vph_{00}^2}{\ol{M}_H^2}
      }\nnn
    \ds{\hat{L}_3^{(H)}} &=& \ds{2
      \left(Z_\Phi+2X_\Phi^-\vph_{00}^2\right)^2
      \frac{\vph_{00}^2}{\ol{M}_H^2}
      }\; .
  \end{array}
\end{equation}

Finally, one may also extract the interactions between the
pseudoscalar Goldstone bosons and the pseudoscalar singlet $\vth$ to
linear order in $\vth$. Taking the singlet--octet mixing into account
they provide the vertices which are needed for a computation of the
decay widths for $\eta^\prime\ra\eta\pi\pi$ and
$\eta^\prime\ra3\pi_0$. Beyond those terms arising from the implicit
$\vth$--dependence of $\Lc_{\chi PT}^{(0)}+\Lc_{\chi PT}^{(1)}$,
(\ref{TTT05}), (\ref{BBB57}), there are also explicit
$\vth$--dependencies appearing in (\ref{BBB07}). They are given to
linear order in $\vth$ by
\begin{equation}
  \label{VVV02}
  \begin{array}{rcl}
    \ds{\Lc_{\chi PT,\vth}^{(1)}} &=& \ds{
      -i U_\Phi\frac{\vph_{00}^2}{\ol{M}_H^2}\vth\Bigg\{
      \Tr\left[\prl_\mu U\prl^\mu U^\dagger
      \left( U\jmath^\dagger-\jmath U^\dagger\right)\right]}\nnn
    &-& \ds{
      4\vph_{00}\left( Z_\Phi+2X_\Phi^-\vph_{00}^2\right)
      \Tr\left[\prl_\mu\left(\prl_\nu U\prl^\nu U^\dagger\right)
      U\prl^\mu U^\dagger\right]\Bigg\} }\; .
  \end{array}
\end{equation}

\sect{Integrating out the $\eta^\prime$ meson}
\label{IntegratingOutTheEetaPrime}

If one is only interested in the effective interactions of the
pseudoscalar octet mesons one may integrate out also the field $\vth$
which corresponds dominantly to the $\eta^\prime$ particle. The
procedure parallels the one of the preceding section. The mass term
$\ol{M}_\vth^2$ is given by (\ref{EtapMassTerm}), and the linear terms
involving $\tilde{U}$, i.e. $-A^{(\vth)}[\tilde{U}]\vth$ give rise to
effective contributions to the low energy constants for $\tilde{U}$
\begin{equation}
  \label{BBB64}
  \Delta\Lc_{\chi PT}^{(1,\vth)}=
  -\frac{1}{2\ol{M}_\vth^2}
  \left( A_\jmath^{(\vth)}+A_{\rm kin}^{(\vth)}\right)^2\; .
\end{equation}
For $\vth$ integrated out and $U\equiv\tilde{U}$ we denote the
constants appearing in the analog of (\ref{BBB57}) by $\tilde{L}_i$.
The difference between the $\tilde{L}_i$ and $\hat{L}_i$ can be computed
from (\ref{BBB64}) with
\begin{equation}
  \label{BBB65}
  \begin{array}{rcl}
    \ds{A_\jmath^{(\vth)}} &=& \ds{
      \frac{i}{6}\vph_{00}\Tr
      \left(\tilde{U}^\dagger\jmath-\jmath^\dagger\tilde{U}\right)}\nnn
    \ds{A_{\rm kin}^{(\vth)}} &=& \ds{
      0}\; .
  \end{array}
\end{equation}
Here we have omitted terms resulting from $\Lc_{\chi
  PT}^{(1)}+\Lc_{\chi PT,\vth}^{(1)}$
which do not contribute to the $\tilde{L}_i$ since they are of higher
order in the quark mass expansion. One therefore finds
\begin{equation}
  \label{BBB67}
  \Delta\Lc_{\chi PT}^{(1,\vth)}=
  \frac{\vph_{00}^2}{72\ol{M}_\vth^2}
  \left[\Tr\left(
    \tilde{U}^\dagger\jmath-\jmath^\dagger\tilde{U}
    \right)\right]^2\; .
\end{equation}
The dominant contribution of this term is an effective additional mass
term for the $\eta$ meson which is due to $\eta$--$\eta^\prime$
mixing. Equation (\ref{BBB67}) yields
\begin{equation}
  \label{BBB66}
  \tilde{L}_7=-\frac{1}{18}
  \left(Z_\Phi+U_\Phi\vph_{00}+X_\Phi^-\vph_{00}^2\right)^2
  \frac{\vph_{00}^4}{\ol{M}_\vth^2}\; .
\end{equation}
For the remaining $\tilde{L}_i$ there is no contribution from
pseudoscalar singlet exchange such that $\tilde{L}_i=\hat{L}_i$ for
$i\neq7$.

\sect{Estimates of the low energy constants}
\label{LiValuesSection}

Having established in the preceding sections the formal correspondence
between the parameters of the linear meson model and the effective
couplings $\tilde{L}_i$ of the nonlinear sigma model we may estimate
the latter using the phenomenological information about the linear
meson model as input. The experimental input for the linear meson
model considered in \cite{JW96-2} concerns only several
(pseudo--)scalar meson masses and decay constants as well as the
$\eta$--$\eta^\prime$ mixing angle. A comparison with the values of
the low energy constants $L_i$ of chiral perturbation theory (see for
instance \cite{BEG94-1}) therefore implicitly relates different
experimental observations.  We remind the reader at this point that
our values for the low energy constants are effective values which
result in chiral perturbation theory after computing the loop
corrections from pseudoscalar fluctuations. We also note that some of
these constants involve as an important parameter the mass of the
scalar singlet (the $\sigma$ particle) which was not determined by the
analysis of \cite{JW96-2}.

We will start by expressing the $\tilde{L}_i$ determined in the
preceding sections in terms of the parameters introduced in
\cite{JW96-2}.  We can use here the approximations
$\vph_{00}=\ol{\sigma}_0\equiv Z_m^{-1/2}\sigma_0$,
$\nu=Z_m^{-3/2}\ol{\nu}$ which yields
\begin{equation}
  \label{BBB71}
  \begin{array}{rcl}
    \ds{\ol{M}_H^2} &=& \ds{
      2Z_h m_h^2}\nnn
    \ds{\ol{M}_T^2} &=& \ds{
      6Z_s m_s^2}\nnn
    \ds{\ol{M}_\vth^2} &=& \ds{
      \frac{2}{3}\frac{Z_p}{Z_m}
      \sigma_0^2 m_p^{(0)2}=
      \nu\sigma_0^3}\nnn
    \ds{Z_\Phi+U_\Phi\vph_{00}+X_\Phi^-\vph_{00}^2} &=& \ds{
      Z_m}\nnn
    \ds{Z_\Phi+2X_\Phi^-\vph_{00}^2} &=& \ds{
      Z_m\left\{1+\left(\frac{Z_h}{Z_m}\right)^{\frac{1}{2}}
      \omega_m\sigma_0\right\}\equiv Z_m\beta
      }\nnn
    \ds{Z_\Phi+\frac{3}{2}U_\Phi\vph_{00}+
      2X_\Phi^-\vph_{00}^2} &=& \ds{
      Z_m\left\{1+\left(\frac{Z_h}{Z_m}\right)^{\frac{1}{2}}
      \omega_m\sigma_0+\frac{1}{2}
      \left(1-\frac{Z_p}{Z_m}\right)\right\}\equiv Z_m\alpha
      }\; .
  \end{array}
\end{equation}
Here $m_s$ denotes the mass of the sigma resonance and $m_h$ is the
average mass of the scalar octet with $Z_s$ and $Z_h$ denoting the
corresponding wave function renormalization constants, respectively.
The parameters $\alpha$ and $\beta$ depend on the quantity $\omega_m$
\cite{JW96-2} which plays an important role in the
$\eta$--$\eta^\prime$ mixing and can be determined phenomenologically
from the decays $\eta\ra2\gamma$ and $\eta^\prime\ra2\gamma$. 
We obtain
\begin{equation}
  \label{BBB73}
  \begin{array}{rcl}
    \ds{\tilde{L}_1^{(H,T)}} &=& \ds{
      \frac{1}{3}\left(\alpha^2\frac{Z_m}{Z_s}
      \frac{\sigma_0^2}{m_s^2}-
      \beta^2\frac{Z_m}{Z_h}
      \frac{\sigma_0^2}{m_h^2}\right)
      }\nnn
    \ds{\tilde{L}_2^{(H,T)}} &=& \ds{0}\nnn
    \ds{\tilde{L}_3^{(H)}} &=& \ds{
      \beta^2\frac{Z_m}{Z_h}\frac{\sigma_0^2}{m_h^2}
      }\nnn
    \ds{\tilde{L}_4} &=& \ds{
      \frac{1}{3}\left(\alpha\frac{Z_m}{Z_s}
      \frac{\sigma_0^2}{m_s^2}-
      \beta\frac{Z_m}{Z_h}
      \frac{\sigma_0^2}{m_h^2}\right)\equiv
      \frac{1}{3}\alpha\frac{Z_m}{Z_h}
      \frac{\sigma_0^2}{m_h^2}
      \left(\frac{\alpha-\beta}{\alpha}+\gamma\right)
      }\nnn
    \ds{\tilde{L}_5} &=& \ds{
      \beta\frac{Z_m}{Z_h}\frac{\sigma_0^2}{m_h^2}
      }\nnn
    \ds{\tilde{L}_6} &=& \ds{
      \frac{1}{12}\left(\frac{Z_m}{Z_s}
      \frac{\sigma_0^2}{m_s^2}-
      \frac{Z_m}{Z_h}
      \frac{\sigma_0^2}{m_h^2}\right)\equiv
      \frac{1}{12}\frac{Z_m}{Z_h}
      \frac{\sigma_0^2}{m_h^2}\gamma
      }\nnn
    \ds{\tilde{L}_7} &=& \ds{
      -\frac{1}{12}\frac{Z_m}{Z_p}
      \frac{\sigma_0^2}{m_p^{(0)2}}=
      -\frac{1}{18}\frac{\sigma_0}{\nu}
      }\nnn
    \ds{\tilde{L}_8} &=& \ds{
      \frac{1}{4}\frac{Z_m}{Z_h}\frac{\sigma_0^2}{m_h^2}
      }\; .
  \end{array}
\end{equation}

From \cite{JW96-2} we infer the values
\begin{equation}
  \label{BBB74}
  \begin{array}{rcl}
    \ds{\alpha} &=& \ds{0.55\pm0.04}\nnn
    \ds{\beta} &=& \ds{0.53\pm0.02}\nnn
    \ds{\sigma_0} &=& \ds{(53.9\pm0.2)\MeV}\nnn
    \ds{\nu} &=& \ds{
      \left(6410\pm420\right)\MeV}\nnn
    \ds{\left(\frac{Z_h}{Z_m}\right)^{1/2}m_h} &=& \ds{(825\pm10)\MeV}\; .
  \end{array}
\end{equation}
The mass $m_s$ of the $\sigma$ meson as well as the quotient $Z_s/Z_m$
remained undetermined in \cite{JW96-2}. We have parameterized this
uncertainty in (\ref{BBB73}) with the scalar singlet--octet splitting
\begin{equation}
  \label{BBB100}
  \gamma\equiv\frac{Z_h}{Z_s}
  \frac{m_h^2}{m_s^2}-1
\end{equation}
such that presumably $\abs{\gamma}<1$. (In fact, to leading order in
the large--$N_c$ expansion one has exactly $\gamma=0$.)  This leads to
the results in table \ref{tab1} where we have omitted our numerical
results for $\tilde{L}_1^{(H,T)}$, $\tilde{L}_2^{(H,T)}$ and
$\tilde{L}_3^{(H)}$ since they are incomplete without the
contributions from higher derivative terms in the linear meson model.
\begin{table}
\begin{center}
\begin{tabular}{|c||c|c|c|c|c|} \hline
    &
  $\tilde{L}_4\cdot10^{3}$ &
  $\tilde{L}_5\cdot10^{3}$ &
  $\stackrel{ }{\tilde{L}_6\cdot10^{3}}$ &
  $\tilde{L}_7\cdot10^{3}$ &
  $\tilde{L}_8\cdot10^{3}$
  \\[0.5mm] \hline\hline
  (a) &
  $(0.78\pm0.06)(\gamma+0.04)$ &
  $2.26\pm0.10$ &
  $(0.36\pm0.01)\gamma $ &
  $-0.47\pm0.03$ &
  $1.07\pm0.03$
  \\ \hline
  (b) &
  $0.2\pm0.5$ &
  $3.0\pm0.5$ &
  $0.1\pm0.3$ &
  $-0.4\pm0.15$ &
  $1.3\pm0.3$
  \\ \hline
  (c) &
  $-0.1\pm0.5$ &
  $2.0\pm0.5$ &
  $-0.1\pm0.3$ &
  $-0.4\pm0.15$ &
  $1.1\pm0.3$
  \\ \hline
\end{tabular}
\caption{\footnotesize This table shows our estimates of some of the
  $\tilde{L}_i$ in line (a) in comparison with the phenomenological
  results for $L_i(\mu)$ taken at normalization scales $\mu=400\MeV$
  (b) and $\mu=600\MeV$ (c).}
\label{tab1}
\end{center}
\end{table}
For comparison the table also gives recent phenomenological estimates
\cite{BEG94-1} of chiral perturbation theory. The $\tilde{L}_i$ are
here compared with the couplings $L_i(\mu)$ of the nonlinear sigma
model normalized at scales $\mu=400\MeV$ (line $(b)$) and
$\mu=600\MeV$ (line $(c)$). The overall agreement of the linear meson
model estimates with these values is striking. In particular, we may
use the low energy constants $L_5$ and $L_8$ for an estimate of
$\beta$ in chiral perturbation theory:
\begin{equation}
  \label{BBB110}
  \beta_{\chi PT}=\frac{1}{4}
  \frac{L_5}{L_8}=0.5\pm0.3\; .
\end{equation}
This agrees well with the value given in (\ref{BBB74}) which was
extracted in \cite{JW96-2} essentially from the ratio of decay rates
$\eta\ra2\gamma$ vs.~$\eta^\prime\ra2\gamma$.  We emphasize that the
difference of $\alpha$ and $\beta$ from one is entirely due to the
non--minimal kinetic terms in the linear meson model. A linear model
that only includes renormalizable interactions would be in very poor
agreement with phenomenology! The difference between our values for
$\tilde{L}_5$ and $\tilde{L}_8$ and the central ones for $L_5(\mu)$
and $L_8(\mu)$ in line $(b)$ of table \ref{tab1}, respectively,
extracted from \cite{BEG94-1} could easily be absorbed by a somewhat
higher choice of the renormalization scale $\mu$ of chiral
perturbation theory as demonstrated in line $(c)$ of table \ref{tab1}.
On the other hand, relatively large higher order quark mass effects
were observed in the computation of $f_K-f_\pi$ in the linear meson
model \cite{JW96-2} and these effects are not included in chiral
perturbation theory.

It should be noted that the errors of the $\tilde{L}_i$ given in line
(a) of table \ref{tab1} only reflect uncertainties in the
determination of the parameters (\ref{BBB74}) within the linear meson
model. They do not represent a systematic error analysis but rather
follow from the scattering of values quoted for a range of realistic
assumptions in the tables of \cite{JW96-2}. We believe, nevertheless,
that these uncertainties are not too far from realistic errors once
the information on higher quark mass corrections and propagator
effects also contained in the linear meson model is properly included.
In fact, the replacement $\vph_{00}\ra\ol{\sigma}_0$ in our
computation of the $\tilde{L}_i$ already includes resummed higher
order contributions from the singlet mass term $(m_u+m_d+m_s)/3$.

Our result for the scalar exchange contribution
$\tilde{L}_3^{(H)}=(1.29\pm0.14)\cdot10^{-3}$ has opposite sign and is
much smaller than the value $L_3=-(4.4\pm2.5)\cdot10^{-3}$ given in
\cite{BEG94-1}.  We take this as a strong indication that kinetic
terms with four derivatives in the linear meson model are dominant as
suggested by vector meson exchange dominance \cite{EGPR89-1}.  This is
further substantiated by the fact that there is no contribution to
$\tilde{L}_2$ due to scalar meson or $\eta^\prime$ exchange in
contrast to the non--vanishing value $L_2=(1.7\pm0.7)\cdot10^{-3}$
obtained from $D$--wave $\pi\pi$ scattering \cite{BEG94-1}.
Nevertheless, we wish to note that our results (\ref{BBB73}) for
$\tilde{L}_1^{(H,T)}$, $\tilde{L}_2^{(H,T)}$ and $\tilde{L}_3^{(H)}$,
i.e. for the scalar meson exchange contributions to $L_1$, $L_2$ and
$L_3$, respectively, agree qualitatively with those of
\cite{EGPR89-1}.

\sect{Conclusions}
\label{Conclusions}

Within the linear meson model we have estimated the effective
couplings $\tilde{L}_4$--$\tilde{L}_8$ of the non--linear sigma model
for the light pseudoscalar mesons. They correspond to effective
vertices related to interactions appearing in next to leading order
chiral perturbation theory. Whereas the low energy couplings
$L_4$--$L_8$ of chiral perturbation theory are the renormalization
scale (and scheme) dependent couplings of a perturbative expansion,
the effective couplings $\tilde{L}_i$ already include all quantum
fluctuations and correspond to $1PI$ vertices. They are independent of
the renormalization scale $\mu$ but involve typical momenta $q$ of the
fields appearing in the $n$--point functions. We evaluate the
effective kinetic terms at the ``pole'' for an average pseudoscalar
octet mass, i.e., $q_0^2=-(2M_{K^\pm}^2+M_{\pi^\pm}^2)/3$. A
comparison with the low energy constants $L_i(\mu)$ for $\mu^2=-q_0^2$
shows good agreement within errors (see table~\ref{tab1}). The central
values almost coincide for $\mu\simeq600\MeV$. A computation of the
$1PI$ vertices $\tilde{L}_i$ within chiral perturbation theory would
be very valuable for a more precise connection between the linear
meson model and chiral perturbation theory. In view of the
comparatively small uncertainties for the $\tilde{L}_i$ this may lead
to a more precise determination of some of the $L_i$.

In our version of the linear meson model the explicit chiral symmetry
breaking due to current quark masses appears only in form of a linear
source term. In this approximation the effective couplings
$\tilde{L}_4$--$\tilde{L}_8$ are entirely determined by the exchange
of scalar $0^{++}$ octet and singlet states. The hypothesis that these
couplings are dominated by scalar exchange has been found earlier
\cite{EGPR89-1} to be in agreement with observation. There
large--$N_c$ results were employed to estimate $L_4$, $L_6$ and $L_7$
whereas $L_5$ and $L_8$ were determined from phenomenological input.
We provide here a quantitative framework for a computation of these
couplings based on phenomenological estimates of the couplings in the
linear meson model. At the present stage the couplings $\tilde{L}_4$
and $\tilde{L}_6$ still depend on the unknown scalar octet--singlet
splitting $\gamma$, (\ref{BBB100}). It would be very interesting to
determine this parameter from phenomenological considerations within
the linear meson model. Furthermore there is the prospect that many of
the parameters of the linear meson model are actually determined by a
partial infrared fixed point behavior \cite{JW96-1} of the running
couplings in the linear quark meson model.  This would lead to a
theoretical ``prediction'' for some of the low energy constants
$\tilde{L}_i$ independent of phenomenological input. For the time
being $\tilde{L}_5$, $\tilde{L}_7$ and $\tilde{L}_8$ are determined
effectively in terms of masses and mixing in the $\eta$--$\eta^\prime$
system and the difference $f_K-f_\pi$. The ratio
$\tilde{L}_4/\tilde{L}_6\simeq2$ follows without further input
parameter.

\vspace{1cm}
\noindent{\bf Acknowledgment}: C.W.~would like to thank H.~Leutwyler
for discussions.
\vspace{0.5cm}


\end{document}